\documentclass{aa}

\usepackage{graphicx}
\usepackage{natbib}
\usepackage{scalerel}

\usepackage[table]{xcolor}

\usepackage[nolist,nohyperlinks]{acronym}

\usepackage{txfonts}
\usepackage[pdfencoding=auto,psdextra]{hyperref}
\hypersetup{
    breaklinks=true,
    colorlinks=true,
    linkcolor=blue,
    filecolor=magenta,      
    urlcolor=blue,
    citecolor=blue
}
\makeatletter
\renewcommand*\aa@pageof{, page \thepage{} of \pageref*{LastPage}}
\makeatother

\usepackage[utf8]{inputenc}

\renewcommand{\linenumbers}[0]{}

\usepackage{euclid}

\usepackage[nameinlink,capitalise]{cleveref}
\crefname{section}{Sect.}{Sects.}
\Crefname{section}{Section}{Sections}
\crefname{figure}{Fig.}{Figs.}
\Crefname{figure}{Figure}{Figures}
\crefname{equation}{Eq.}{Eqs.}
\Crefname{equation}{Equation}{Equations}
\crefname{table}{Table}{Tables}
\crefname{appendix}{Appendix}{Appendices}
\Crefname{appendix}{Appendix}{Appendices}

\usepackage[normalem]{ulem} 

\newcommand{\pz}{\phantom{0}}
\newcommand{\po}{\phantom{1}}

\begin{document}
%
%
\acrodef{AEGIS}{All-Wavelength Extended Groth Strip International Survey}
\acrodef{CDFS}{Chandra Deep Field South}
\acrodef{COSMOS}{Cosmic Evolution Survey}
\acrodef{DES}{Dark Energy Survey}
\acrodef{DESI}{Dark Energy Spectroscopic Instrument}
\acrodef{DPDD}{Data Product Description Document}
\acrodef{DR1}{First Data Release}
\acrodef{EAF}{Euclid Auxiliary Field}
\acrodef{EC}{Euclid Consortium}
\acrodef{EDF}{Euclid Deep Field}
\acrodef{EDS}{Euclid Deep Survey}
\acrodef{EMDS}{Euclid Medium-Deep Survey}
\acrodef{EUDF}{Euclid Ultra-Deep Field}
\acrodef{ESA}{European Space Agency}
\acrodef{ERO}{Early Release Observations}
\acrodef{EWS}{Euclid Wide Survey}
\acrodef{EXT}{EXTernal data}
\acrodef{FGS}{Fine Guidance Sensor}
\acrodef{FoV}{field of view}
\acrodef{GC}{galaxy clustering}
\acrodef{HKTM}{housekeeping and telemetry}
\acrodef{ILS}{Independent Legacy Scientist}
\acrodef{LE1}{Level 1}
\acrodef{LE2}{Level 2}
\acrodef{LE3}{Level 3}
\acrodef{MOC}{Mission Operation Centre}
\acrodef{NIR}{near-infrared}
\acrodef{NISP}{Near-Infrared Spectrometer and Photometer}
\acrodef{OU}{Organisation Unit}
\acrodef{Pan-STARRS}{Panchromatic Survey Telescope and Rapid Response System}
\acrodef{PDC}{phase-diversity calibration}
\acrodef{PF}{processing function}
\acrodef{PHZ}{photometric redshift}
\acrodef{PSF}{point-spread function}
\acrodef{PV}{performance-verification}
\acrodef{Q1}{First Quick Data Release}
\acrodef{ROS}{Reference Observation Sequence}
\acrodef{RSD}{Reference Survey Definition}
\acrodef{SAS}{Science Archive System}
\acrodef{SED}{spectral energy distribution}
\acrodef{SGS}{Science Ground Segment}
\acrodef{SHE}{SHEar measurements}
\acrodef{SIR}{Spectroscopy InfraRed}
\acrodef{SPE}{NISP SPEctra analysis}
\acrodef{UNIONS}{Ultraviolet Near-Infrared Optical Northern Survey}
\acrodef{VIS}{VISible instrument}
\acrodef{WAVES}{Wide-Area VISTA Extragalactic Survey}
\acrodef{WL}{weak lensing}

\title{Euclid First Data Release 1 (DR1)}
\subtitle{Data release overview}



\author{\normalsize Euclid Collaboration: H.~Aussel$^{1}$\thanks{\email{herve.aussel@cea.fr}} et al.}

\institute{$^{1}$  Université Paris-Saclay, Université Paris Cité, CEA, CNRS, AIM, 91191 Gif-sur-Yvette, France}

%
%
%
%


%
%
\abstract{
This paper describes the contents of the Euclid First Data Release (DR1), together with the set of papers that document the data-processing and the data products. DR1 consists of two parts: first `DR1-Foundation', releasing images, spectra, and catalogues in November 2026. Second, the shear measurements and the `Level~3' products required for cosmological analyses will be added in June 2027 to complete the full DR1. This data set contains around a billion catalogued objects, mainly from the Euclid Wide Survey (1975\,deg$^2$) and the Euclid Deep Fields (77\,deg$^2$).
}
%
%
%
%
   \titlerunning{Euclid Data Release 1 (DR1): Overview}
   \authorrunning{Euclid Collaboration: H.~Aussel et al.}
   
   \maketitle
%
%
%
%
   
\section{\label{sc:Intro}Introduction}

\Euclid is a space mission of the \ac{ESA} with contributions from the National Aeronautics and Space Administration, with the primary goal of studying dark matter and dark energy using two main probes, weak gravitational lensing and galaxy clustering \citep{Laureijs11}. \Euclid's 1.2-m primary mirror has a \ac{FoV} of 0.54\,deg$^2$, with the optics designed to conduct the \ac{EWS}, covering 14\,000\,deg$^2$ of the extragalactic sky \citep{Scaramella-EP1}, along with the smaller and deeper \ac{EDS}.  
\Euclid contains two instruments, the \ac{VIS} and the \ac{NISP}, which are described in
\citet{EuclidSkyVIS} and \citet{EuclidSkyNISP}, respectively.
\ac{VIS} is a broad-band optical imager with a spatial resolution of \ang{;;0.18} in the \IE band, while \ac{NISP} combines the capabilities of an imager in the \ac{NIR} bands \YE, \JE, and \HE \citep{Schirmer-EP18} with a near-infrared slitless spectrograph \citep{EP-Gillard}. Further details of the \Euclid mission can be found in the earlier overviews \citet{EuclidSkyOverview} and \citet{Q1-TP001}, and the more technical papers referred to in them.
Since much of the content of these papers are also relevant to DR1, here we concentrate on the details that differ for this new release.

Although conceived as a cosmology mission, \Euclid's wide \ac{FoV} and deep imaging capabilities in the optical and \ac{NIR}, along with the complementary ground-based imaging, make its data set useful for studying a wide range of astrophysical topics, as evidenced by the large number of publications that have already appeared. The first data release from \Euclid was the \ac{ERO} programme \citep{EROData}, highlighting just 24 hours of early data.  This was followed by the first `quick' data release, Q1 \citep{Q1-TP001}, containing the \ac{EDS} fields at the depth of the \ac{EWS}, and then Q2 \citep{Q2-TP002}, a survey of the Galactic bulge.

We present here an overview of the first of the major planned data releases from \Euclid, the \ac{DR1}.  The initial part of \ac{DR1}, known as DR1-Foundation \citep{DR1cite}, contains imaging, spectroscopy, and catalogues from a collection of fields of varying depths covering more than 2000\,deg$^2$. The full set of DR1 products, including weak lensing and 3-dimensional galaxy catalogues, together with correlation functions and cosmological analyses from these same areas of the sky will follow in June 2027.  In \cref{sc:DR1F}, we present the definition of DR1-Foundation with respect to the full DR1. In \cref{sc:dataproc}, we summarise the data processing and the available data products and in \cref{sc:datapapers}, we give abbreviated descriptions of the other papers supporting DR1.

\section{\label{sc:DR1F}The DR1-Foundation data release}

The \Euclid mission will deliver three major data releases, the first one corresponding to approximately the first year of the \Euclid survey (DR1), the second to the first three years of the survey (DR2), and finally the complete survey (DR3). Here we describe what is contained in DR1-Foundation in the context of \ac{DR1}. The \ac{DR1} content is defined in \citet{Euclid:DR1Def}. We review here the main points of this memorandum, which is necessary since \ac{DR1} is not limited to the areas of the main \Euclid surveys, but includes several other regions of the sky.

The \Euclid survey did not start directly after the satellite's launch, so there were earlier periods, including what is known as the \ac{PV} phase, as well as the \ac{PDC} campaign. \ac{DR1} also contains calibration observations obtained during or before the routine survey phase, together with science observations that were obtained outside the survey. We briefly describe the timeline of the mission below.

\subsection{\label{ssc:history}The phases of the \Euclid mission} 
The \Euclid mission so far can be split into four phases: (1) launch and commissioning; (2) \ac{PV}; and (3) the \ac{PDC} campaign; and (4) the \Euclid survey. These are described in detail in \citet{DR1-TP027}, and we review them briefly here.  

\Euclid was launched on 1 July 2023, and travelled to its nominal halo orbit around the Earth-Sun second Lagrange point (L2) during the month of July \citep[see figure 1 of][for details]{DR1-TP026b}. During this time, the spacecraft was thermally decontaminated, the platform and instruments were switched on, and commissioning ran until 31 July. 

\ac{PV} was scheduled for 60 days, beginning on 1 August 2023. Due to insufficient tracking stability of the \ac{FGS}, it was interrupted several times,
including replanning of the \ac{PV} to advance observations with lower demands on tracking stability; these observations were interleaved with pointing-stability tests. The \ac{FGS} was successfully recommissioned during 23--30 September 2023, including visits of the self-calibration field, the 
\ac{ERO} \citep{EROData}, and a dust-obscured area in the Orion Molecular Cloud that was released with the \acl{Q1} \citep[\acs{Q1};][]{Q1-TP001}\acused{Q1}. \ac{PV} transitioned around 1 December 2023 into the \ac{PDC} campaign to characterise \Euclid's wavefront. The observations of stellar calibration fields -- using intrafocal, focal, and extrafocal settings -- were separated by approximately 7 days during which the satellite pointing was constrained in order to maintain high thermal stability. Details about the calibration and survey-related aspects of \ac{PV} and \ac{PDC} are given in \cite{DR1-TP026} and \cite{DR1-TP027}. 

The \Euclid survey proper started on 14 February 2024. It consists of observations of the \ac{EWS}, the \ac{EDS}, and the \acp{EAF}, interleaved with various on- and off-sky calibration observations. The last observation of the nominal survey is currently scheduled for 2 January 2030. 

Ice contamination in the instrument cavity led to a gradual throughput loss for the VIS instrument, compromising the mission objectives. The first decontamination procedure on 14 March 2024 fully restored the throughput, which then began to rapidly decline at an accelerating rate until 8 June 2024, when \Euclid was decontaminated once more. This `ice age' is discussed in sections 5 and 6 of \citet[]{DR1-TP026b}. The VIS throughput has been remarkably stable since, gradually decreasing by 1--1.5\% over the following 2.5 years. The NISP detectors accumulate ice, leading to a slow wavelength-dependent evolution of sensitivity that is well captured by calibrations, but in DR1 not yet corrected for the blue-grism data.

\subsection{\label{ssc:DR1Def}DR1 definition}

\ac{DR1} is defined loosely as the observations that were obtained since launch, up to the end of the first year of the \ac{EWS}. To be included, these observations must be `complete' and `valid'. `Complete' means that, in the case of the \Euclid survey, the ancillary ground-based data (referred to as `EXT') needed for the scientific exploitation of the observations must be available to the \ac{SGS}. `Valid' means that the data quality meets a series of technical requirements. Exposures can be invalidated by external and internal factors, such as solar radiation storms, or guiding and instrument errors. The validity flags set by the instrument processing functions and initial data-merging steps are explained in the articles describing the \acl{PF} (\acs{PF}; see below)\acused{PF}. 
The spacecraft's technical reliability is very high, with more than 99\% of all pre-planned scientific and telemetry files retrieved on ground \citep[see section 2 in][]{DR1-TP026b}. 

The ice age occupied around three months early in the \ac{EWS}. This period was excluded from DR1 because of considerable uncertainties in the photometric and \ac{PSF} calibration of the VIS data. DR1 
corresponds to approximately 12 months of survey observations outside the ice age, concluded by the last observations of the self-calibration block around 2025-05-21T19:12:00Z.

\begin{figure*}
\centering
\includegraphics[width=1\textwidth]{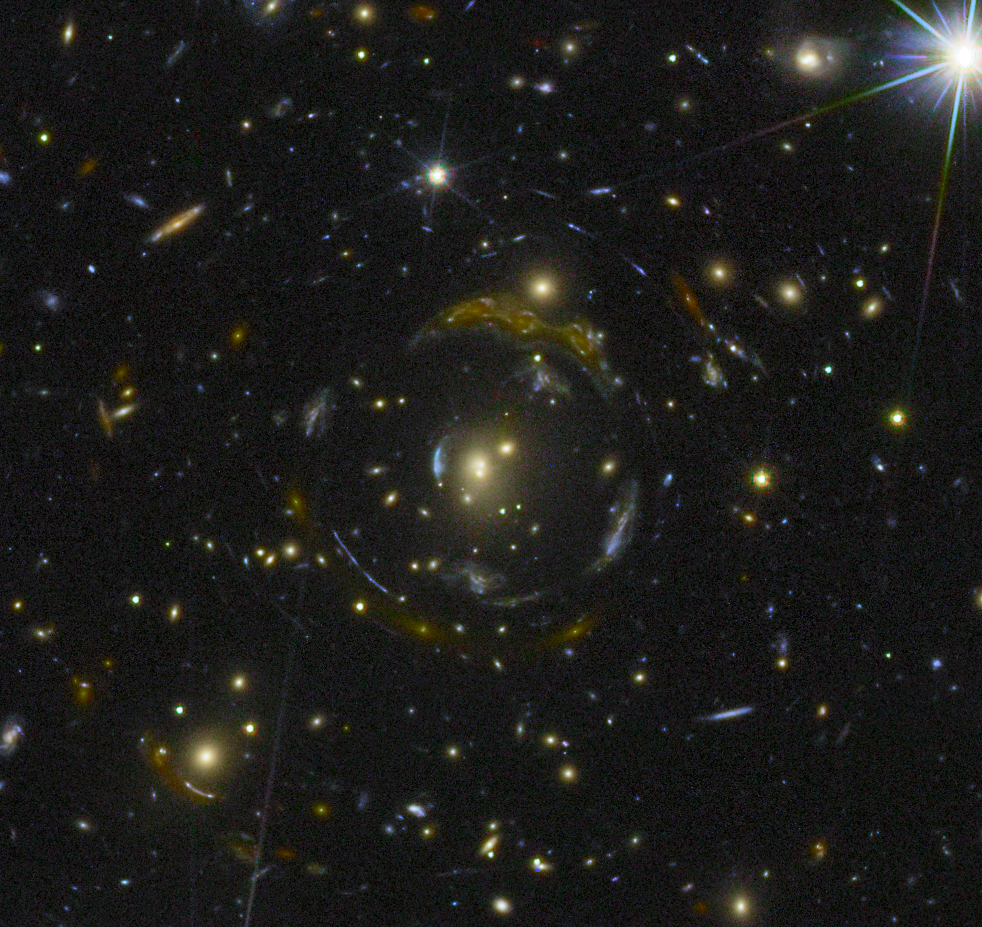}
\caption{\label{fig:carousel}The Carousel Lens at redshift $z=0.49$, built from DR1 \ac{EWS} data. The image is $\ang{;1.5}\times\ang{;1.4;}$ across, with north up and east to the left. Strong lensing has been one of the science highlights coming from DR1-Foundation.}
\end{figure*}

\subsection{Data-set definitions}\label{ssc:datasetdefinitions}

The data released in \ac{DR1} is not a single homogeneous data set. The raw frames, or \ac{LE1} products,\footnote{See section 7.3.1 of \citet{Laureijs11} for the definition of the levels in the \Euclid data processing system.} for all valid and complete observations are distributed through the \ac{SAS},\footnote{The entry point is the Euclid Data Space:\\\url{https://euclid.dataspace.esa.int}} comprising both calibration and survey observations. However, some of the calibration observations do not produce \ac{SGS}
products. Similarly, not all observations are processed in the same way. For example, individual visits to the \acp{EDF} will be processed as \ac{EWS} observations, but they will also be co-added to produce deep data sets. 

\begin{table}
\caption{\label{tab:datasets} Definition of DR1 data sets.}
\tabcolsep=0.18cm 
\begin{tabular}{lp{7cm}}
\hline\hline
\noalign{\vskip 2pt}
Data set & Definition \\
\hline
\noalign{\vskip 2pt}
\ac{EWS} & CALBLOCK-F-006  \\
\ac{EDS} & CALBLOCK-F-002 and CALBLOCK-F-003  \\
\acp{EAF} & CALBLOCK-F-004, CALBLOCK-F-005 and PV Observation ID 65879 \\
SelfCal & CALBLOCK-F-001 except Patch ID $91^{a}$ and PV Observation IDs 65765, 65776, 65803, 65811 \\
PSF & CALBLOCK-F-008 except Patch ID $158^{a}$ \\
F-010 & CALBLOCK-F-010 \\
F-014 & CALBLOCK-F-014 \\
ILS & Observation IDs 65914 to 66048 (135 Obs.\ IDs)\\
WAVES & Observation IDs 66059 to 66190 (132 Obs.\ IDs)\\
SSO & Observation IDs 66202 to 66350 (149 Obs.\ IDs) \\
EMDS-1 &  Observation IDs 66358 to 66491 (134 Obs.\ IDs)\\
EMDS-2 & Observation IDs 66517 to 66660 (129 Obs.\ IDs) \\
EMDS-3 & Observation IDs 66680 to 66819 (136 Obs.\ IDs) \\
EMDS-4 & Observation IDs 66826 to 66962 (136 Obs.\ IDs)  \\
EMDS & Combination of EMDS-1 to EMDS-4 \\
Other & All other CALBLOCKs \\
\hline
\end{tabular}
 \footnotesize
 \tablefoot{$^{(a)}$ These patches were lost to high solar activity. See \cref{ssc:datasetdefinitions} for brief descriptions of these data sets.}
\end{table}

Activities performed during an observation are defined by their `ObservationType', whose name is either CALBLOCK-PV-XXX for the PV phase or CALBLOCK-F-XXX for the routine phase \citep[see][for a full list]{DR1-TP026}. Observations are identified by their Observation~ID. If they are, for example, part of the \acl{ROS} \citep[\acs{ROS};][]{DR1-TP027}\acused{ROS} executed at every survey field, then they all share the same Observation~ID. Groups of survey fields taken contiguously in time and on sky form a `survey patch', and share its Patch~ID. These labels and identifiers uniquely identify the content of DR1 
(\cref{tab:datasets}).

The sky content of DR1 is presented in \cite{DR1-TP027}. The largest constituents are two irregularly shaped blocks, EWS-South and EWS-North, covering approximately 1600\,deg$^2$ and 500\,deg$^2$, respectively. The \ac{EDS} fields have gained considerable depth since their first release in Q1 \citep{Q1-TP001}: The EDF-N, EDF-S, and EDF-F were visited 11, 10, and 2 times, respectively, each visit (or `pass') having the same depth as a single \ac{EWS} field. The red grism was used throughout, apart from three passes of the EDF-N, for which the \ac{ROS} was configured with the blue grism. The final effective exposure-time ratio is set to around 25:15 (blue:red) visits, adjusted for zodiacal background brightness. 

The \ac{EMDS} is around 1\,mag deeper and one order of magnitude smaller than the \ac{EWS}, and consists of the four sub-surveys EMDS-1 to EMDS-4. The EMDS-1 and EMDS-2 parts were taken during the \ac{PDC} thermal stabilisation periods, which required NISP in idling mode to maximise thermal stability; only VIS data are available for these. The \ac{EMDS} is part of DR1-Foundation.

The \acp{EAF} consist of well-known extragalactic fields used for calibration of photometric redshifts, and to assess the effect of colour gradients within galaxies on shear measurements in the presence of a chromatic \ac{PSF}.  \citep{DR1-TP011}. For DR1, these are the \acl{COSMOS} \citep[\acs{COSMOS};][]{Scoville:2007a}\acused{COSMOS}, \acl{CDFS} \citep[\acs{CDFS};][]{Giacconi:2002}\acused{CDFS}, and \acl{AEGIS} \citep[\acs{AEGIS};][]{Davis:2007}\acused{AEGIS} fields. Others will be observed later during the survey.  

The self-calibration (or `SelfCal') field covers around 2.0\,deg$^2$ near the north ecliptic pole (NEP), within \Euclid's continuous viewing zone, and is within the area of the EDF-N. Data were acquired through CALBLOCK-F-001 and CALBLOCK-PV-001, covering the full area.  More data have been acquired through CALBLOCK-F-014, which executes the \ac{ROS} at the centre of the self-calibration field for throughput monitoring during \ac{PDC} and after a thermal decontamination of the spacecraft. Together, these data comprise the extremely deep \ac{EUDF}, which will be released early in 2027 (McCracken et al., in prep.).

Several PSF-calibration fields (CALBLOCK-F-008) are also part of DR1. They are observed every 2--6 weeks to obtain representative \ac{PSF} measurements for every survey patch. A special pattern of 100 dithers is used, at each of which VIS takes a 289-s exposure. During this time, NISP takes off-sky nonlinearity calibrations. PSF-calibration fields are observed at least once also with one \ac{ROS} equivalent, to secure NISP data for stellar classification. 

The `ILS' fields were defined by the \acp{ILS} and `WAVES' is part of the Wide-Area VISTA Extragalactic Survey \citep{Driver:2019}. Lastly, `SSO' is a survey field designed for Solar System objects, close to the ecliptic plane. These three components are described in more depth in~\citet{DR1-TP027}.

\subsection{\label{ssc:dr1fdef}DR1-Foundation}

All \ac{DR1} data were reprocessed with the \ac{SGS} processing function in its frozen DR1 configuration, ensuring maximum homogeneity (\cref{sc:dataproc}). Reprocessing started in May 2025 for the \acl{EXT} \citep[\acs{EXT};][]{DR1-TP006}\acused{EXT}. On the spectroscopic side, the \ac{LE2} processing was completed at the end of September 2025. On the imaging side, the \ac{LE2} products up to PHZ were available mid-September 2025, but it took until the end of October 2025 to produce the \ac{SHE} \ac{PSF} model and shape measurements. The full reprocessing timeline is given in figure 8 of \citet{DR1-TP033}. 

Once the products have been computed, they undergo a series of tests by the relevant \ac{OU} to check whether they are compliant with the mission's scientific requirements. This validation can take up to 1--2 months in the case of more complex products. In early 2026, it became evident that the \ac{PSF} model underestimated the size of the VIS \ac{PSF} and would thus compromise the shear analysis. 

With \ac{SHE} processing requiring at least 2 months, this would have left less than 6 months for a full cosmological analysis before the nominal release date of 12 November 2026. Since all other \ac{LE2} products had been internally released a year before this date, a staggered approach was decided for the public release. \ac{DR1}-Foundation occurs on 12 November 2026, including LE1 and validated LE2 data products, such as images, spectra, and catalogues. The LE3 data products, comprising shape measurements, \ac{WL} and \ac{GC} correlation functions, and the cosmological analyses, are added in June 2027. Together, they form the full DR1; the data products that comprise DR1-Foundation will not be re-released in June 2027.

\begin{table*}
\caption{\label{tab:le1}   \ac{LE1} Data products released as part of DR1-Foundation, listing the number of LE1 products of each type for each data set.}
\begin{center}
\begin{tabular}{lc@{\hspace{1.0cm}}cc@{\hspace{1.0cm}}cccc}
\hline\hline
\noalign{\vskip 2pt}
  & Scheduled          & \multicolumn{2}{c}{\uline{VIS raw frames}} & \multicolumn{4}{c}{\uline{NISP raw frames}} \\
  \noalign{\vskip 2pt}
Data set & observations & VIS nominal & VIS short & NISP spectro & NISP \YE & NISP \JE & NISP \HE \\
\hline
\noalign{\vskip 2pt}
EWS & 19588 & 17143 & \pz8572 & 17231 & 17231 & 17231 & 17231 \\
EDFs & \pz4568 & \pz4339 & \pz2170 & \pz4392 & \pz4392 & \pz4392 & \pz4392 \\
EAFs & \pz\pz896 & \pz\pz896 & \pz\pz358 & \pz\pz896 & \pz\pz896 & \pz\pz896 & \pz\pz896 \\
SelfCal & \pz1216 & \pz1140 & \pz\pz\pz30 & \pz1140 & \pz1140 & \pz1139 & \pz1139 \\
PSF & \pz2136 & \pz1964 & \pz\pz\pz\pz0 & \pz\pz\pz\pz0 & \pz1648 & \pz\pz\pz\pz0 & \pz\pz\pz\pz0 \\
F014 & \pz\pz104 & \pz\pz\pz88 & \pz\pz\pz32 & \pz\pz\pz89 & \pz\pz\pz88 & \pz\pz\pz88 & \pz\pz\pz88 \\
F010 & \pz\pz856 & \pz\pz819 & \pz\pz\pz22 & \pz\pz856 & \pz\pz\pz16 & \pz\pz\pz16 & \pz\pz\pz16 \\
ILS & \pz\pz540 & \pz\pz536 & \pz\pz268 & \pz\pz498 & \pz\pz498 & \pz\pz498 & \pz\pz498 \\
WAVES & \pz\pz528 & \pz\pz520 & \pz\pz260 & \pz\pz\pz44 & \pz\pz\pz44 & \pz\pz\pz44 & \pz\pz\pz44 \\
SSO & \pz\pz596 & \pz\pz596 & \pz\pz298 & \pz\pz596 & \pz\pz596 & \pz\pz596 & \pz\pz596 \\
EMDS1 & \pz\pz536 & \pz\pz532 & \pz\pz266 & \pz\pz\pz\pz0 & \pz\pz\pz\pz0 & \pz\pz\pz\pz0 & \pz\pz\pz\pz0 \\
EMDS2 & \pz\pz516 & \pz\pz516 & \pz\pz258 & \pz\pz\pz\pz0 & \pz\pz\pz\pz0 & \pz\pz\pz\pz0 & \pz\pz\pz\pz0 \\
EMDS3 & \pz\pz544 & \pz\pz455 & \pz\pz228 & \pz\pz528 & \pz\pz528 & \pz\pz528 & \pz\pz528 \\
EMDS4 & \pz\pz544 & \pz\pz427 & \pz\pz217 & \pz\pz528 & \pz\pz528 & \pz\pz528 & \pz\pz528 \\
\hline
\noalign{\vskip 2pt}
Total & 48524 & 39274 & 13375 & 30764 & 30221 & 28562 & 28562 \\
\hline
\noalign{\vskip 2pt}
Other &  & \multicolumn{2}{c}{\pz\pz9699$^{a}$ }    & \pz3966 & \pz2616 & \pz2606 & \pz2606 \\
\hline
\end{tabular}
 \footnotesize
 \tablefoot{See \cref{ssc:datasetdefinitions} for an explanation of the data set names.   $^{(a)}$ For the `Other` category, the distinction between VIS nominal and short exposures has no meaning because calibrations are obtained with many different exposure times.}
\end{center}
\end{table*}

\section{\label{sc:dataproc}  Data processing and data products }

\begin{figure*}
    \includegraphics[width=\textwidth]{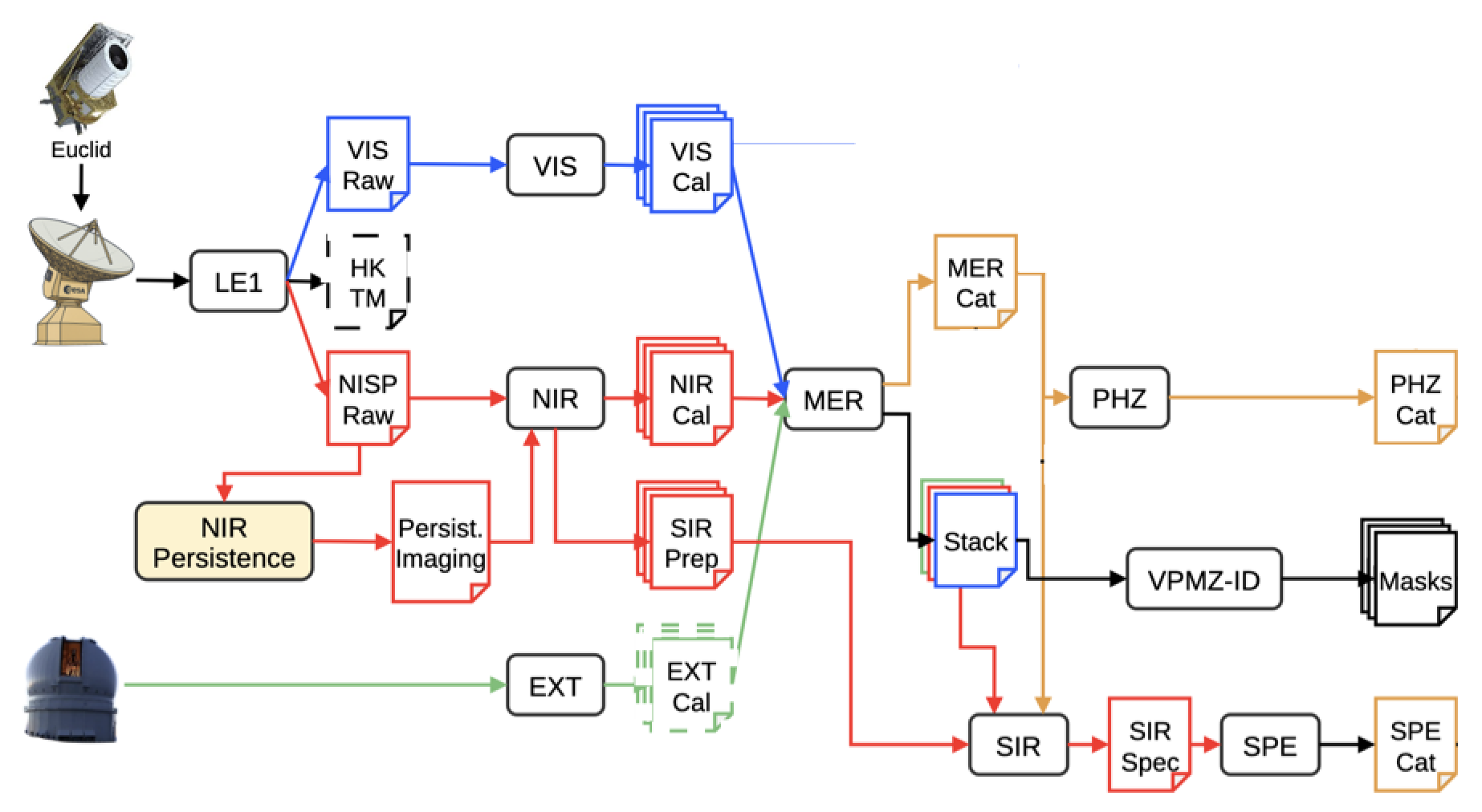}
    \caption{Overview of \ac{DR1} data processing, showing the main data products making up the \ac{DR1} data release, together with the \ac{SGS} \ac{PF} creating them, and outlining the flow of the various data sets. `Raw' products refer to products created from the decommuted telemetry without any further processing. `Cal' products refer to calibrated data, with instrument signatures being removed and data put on a physical scale. `Stack' refers to image combinations, `Spec' to combined 1D spectra, and `Cat' to catalogues. This is an adapted figure~3 from \citet{DR1-TP033} to show the process for DR1-Foundation only. Products indicated with a dashed outline are created and used in the pipeline, but are not distributed as part of DR1-Foundation.}\label{fig:dp}
\end{figure*}

The \Euclid mission data are processed by the \ac{EC} \ac{SGS}. \citet{DR1-TP033} describes the architecture of this distributed computing system, together with account of the DR1 data processing. They provide in their figure 3 an overview of the pipeline with the \ac{PF} and their main products, and the connection between them. This figure has been 
updated in \cref{fig:dp} here to reflect the processing chain and data products distributed as part of DR1-Foundation. This processing chain is very similar to the one executed for the Q1 data release~\citep{Q1-TP001}, but each PF has been substantially improved to produce DR1-Foundation; this is demonstrated in \cref{fig:carousel},\footnote{Image built with \url{http://pypi.org/project/eummy/}} showing the Carousel Lens system \citep{huang2021,odonnel2022}. We now briefly review these improvements and indicate the relevant papers where complete information can be obtained.

\subsection{\label{ssc:le1}Telemetry decommutation and LE1 products}

The LE1 processing is identical to that performed for Q1 and described in~\citet{Q1-TP001}. The LE1 PF produces the VIS 
raw-frame product, the NISP raw-frame product, and the \ac{HKTM} product from the telemetry received from the \ac{MOC}. These are described in detail in the \ac{DPDD}.\footnote{\url{https://euclid.esac.esa.int/dr/dr1/dpdd}} Only VIS and NISP raw frames are distributed with DR1-Foundation. The \ac{HKTM} products are used by the \ac{SHE} \ac{PF} to reconstruct the pointing history during an exposure and will de distributed at the time of the full \ac{DR1}. 

The scheduled observations column of \cref{tab:le1} allows us to estimate the reliability of the \Euclid mission observations. In the \ac{EWS}, for \ac{VIS}, 17\,143 nominal exposures were collected from 19\,588 scheduled in the \ac{RSD}, or about 87\%, within the mission requirements. Missing data or failures are due to, for example, invalid data products or missing ground-based \ac{EXT} coverage data.  This last point is a factor that limited \ac{PSF} calibration for the DR1 processing, because stellar \acp{SED} were not yet completely available. We expect higher completion rates for DR2 and DR3.

\subsection{\label{ssc:VIS}VIS data processing}
\setlength{\tabcolsep}{12pt} 
\begin{table*}
\caption{\label{tab:le2visnisp}Number of VIS and NISP LE2 nominal-exposure validated calibrated frames per data set, and the corresponding areas. }
\begin{center}
    \begin{tabular}{lcc@{\hspace{2.0cm}}cccc}
\hline\hline
\noalign{\vskip 2pt}
    Data set & No. exposures & Area
    & \multicolumn{3}{c}{No. exposures} & Area \\
    & \IE & [deg$^2$]   &      \YE & \JE & \HE &  [deg$^2$] \\
\noalign{\vskip 2pt}
\hline
\noalign{\vskip 2pt}
EWS & 16913 & 2092.2$^{\phantom{a}}$
  & 16111 &  15787 & 15977 & 1975.4$^{\phantom{a}}$ \\
\hline
\noalign{\vskip 2pt}
EDFs & \pz4568 & \pz\pz77.0$^{\phantom{a}}$
  &  \pz3942 &  \pz3898 & \pz3930 & \pz\pz76.8$^{\phantom{a}}$ \\
\hline
\noalign{\vskip 2pt}
EDF-N & \pz2408 & \pz\pz24.3$^{a}$
  &  \pz1806 &  \pz1756 & \pz1799 & \pz\pz19.8$^{a}$ \\
EDF-F & \pz\pz184 & \pz\pz13.1$^{a}$
  & \pz\pz182 &  \pz\pz181 & \pz\pz178 & \pz\pz13.0$^{a}$ \\
EDF-S & \pz1976 & \pz\pz27.8$^{a}$
  & \pz1954 &  \pz1961 & \pz1957 & \pz\pz23.2$^{a}$\\
Total & \pz4568 & \pz\pz65.2$^{a}$
  & \pz3942 &  \pz3898 & \pz3934 & \pz\pz56.0$^{a}$ \\
\hline
\noalign{\vskip 2pt}
EAFs & \pz\pz895 & \pz\pz\pz4.8$^{\phantom{a}}$
  & \pz\pz910 &  \pz\pz908 & \pz\pz907 & \pz\pz\pz4.8$^{\phantom{a}}$ \\
SelfCal & \pz1215 & \pz\pz\pz2.2$^{\phantom{a}}$
  & \pz1972 &  \pz1977 & \pz1978 & \pz\pz\pz2.2$^{\phantom{a}}$ \\
PSF & \pz2130 & \pz\pz\pz6.3$^{\phantom{a}}$
  &  \pz\pz\pz\pz0 &  \pz\pz\pz\pz0 & \pz\pz\pz\pz0 & \pz\pz\pz0.0$^{\phantom{a}}$ \\
F010 & \pz\pz856 & \pz\pz\pz1.8$^{\phantom{a}}$
  & \pz\pz\pz15 & \pz\pz\pz15 & \pz\pz\pz15 & \pz\pz\pz1.8$^{\phantom{a}}$ \\
F014 & \pz\pz103 & \pz\pz\pz1.6$^{\phantom{a}}$
  & \pz\pz\pz57 &  \pz\pz\pz54 & \pz\pz\pz55 & \pz\pz\pz1.6$^{\phantom{a}}$ \\
ILS & \pz\pz538 & \pz\pz37.5$^{\phantom{a}}$
  & \pz\pz466 &  \pz\pz470 & \pz\pz471 & \pz\pz35.9 $^{\phantom{a}}$ \\
WAVES & \pz\pz525 & \pz\pz14.8$^{\phantom{a}}$
  & \pz\pz\pz\pz0 &  \pz\pz\pz\pz0 & \pz\pz\pz\pz0 & \pz\pz\pz0.0$^{\phantom{a}}$ \\
SSO & \pz\pz593 & \pz\pz19.2$^{\phantom{a}}$
  & \pz\pz595 &  \pz\pz598 & \pz\pz587 & \pz\pz18.2$^{\phantom{a}}$ \\
EMDS-1 & \pz\pz533 & \pz\pz15.1$^{\phantom{a}}$
  & \pz\pz\pz\pz0 &  \pz\pz\pz\pz0 & \pz\pz\pz\pz0 & \pz\pz\pz0.0$^{\phantom{a}}$ \\
EMDS-2 & \pz\pz516 & \pz\pz13.0$^{\phantom{a}}$
  & \pz\pz\pz\pz0 &  \pz\pz\pz\pz0 & \pz\pz\pz\pz0 & \pz\pz\pz0.0$^{\phantom{a}}$ \\
EMDS-3 & \pz\pz536 & \pz\pz15.9$^{\phantom{a}}$
  & \pz\pz450 &  \pz\pz438 & \pz\pz442 & \pz\pz15.7$^{\phantom{a}}$ \\
EMDS-4 & \pz\pz541 & \pz\pz13.9$^{\phantom{a}}$
  & \pz\pz396 &  \pz\pz388 & \pz\pz395 & \pz\pz12.2$^{\phantom{a}}$ \\
\hline
\noalign{\vskip 2pt}
Total & 33135 & 2268.7$^{\phantom{a}}$
  & 24068 & 23681 & 23910 & 2268.7$^{\phantom{a}}$ \\
\hline
    \end{tabular}
     \footnotesize
 \tablefoot{$^{(a)}$ For the \acp{EDF} this corresponds to the deep area with maximum overlap.}
    \end{center}
\end{table*}

The VIS instrument data are processed by the VIS \ac{PF}. The processing is described in \citet{Q1-TP002}. Numerous tasks, such as cosmic-ray rejection, photometric calibration, background estimation, coaddition, and the content of the delivered data products were improved for \ac{DR1}. These updates, together with a comprehensive assessment of data quality and an outlook for further improvements, are described in \citet{DR1-TP002}.

\begin{figure*}
\centering
\includegraphics[width=0.875\textwidth]{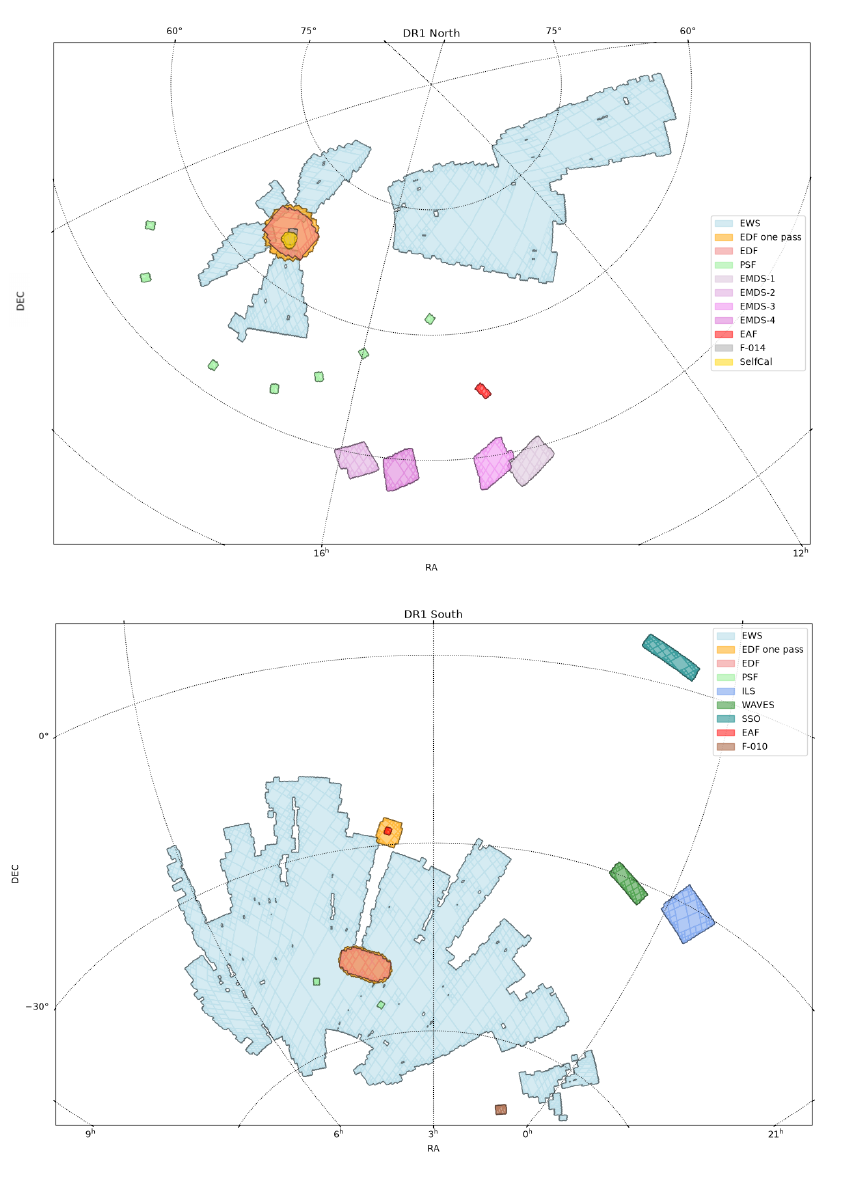}
\caption{\label{fig:visle2}Distribution on the sky of the various DR1 data sets in the north (top) and in the south (bottom). Note that there are no EWS data underneath the EDF-N. In the south, there are no EWS data in the EDF-F, but the area of the EDF-S has been observed as part of the EWS. In the north, the \ac{EAF} area is AEGIS, while in the south, it is the CDFS in the centre of the EDF-F. The third \ac{EAF}, COSMOS, is outside the field of view of these figures.}
\end{figure*}

The number of validated LE2 exposures -- with the nominal integration time of 565\,s, or the 289\,s exposure time for the PSF data set -- are reported in \cref{tab:le2visnisp} for each data set, together with their covered areas. Their distribution on the sky is shown in \cref{fig:visle2}. 

\subsection{\label{ssc:NIR}NIR data processing}

NISP calibrated frames are processed by the NIR \ac{PF} \citep{Q1-TP003}. Substantial improvements were implemented for DR1 \citep{DR1-TP003}, in particular with an updated in-flight calibration plan that allows for better persistence masking. Persistence in the NISP detector is discussed in \citet{DR1-TP038}, where a model predicting the amount of signal is described. 

\Cref{tab:le2visnisp} also gives the number of validated calibrated NIR frames in each NISP photometric bands delivered in DR1-Foundation. The corresponding area where all three bands are valid is indicated in the last column. 

The astrometric calibration of NIR frames is described in \citet{DR1-TP004}. While most of the data are within requirements, a fraction (0.7\%) of the DR1 area has error larger than 50\,mas. A {\tt HEALPix} information map of the astrometric error is provided by the Visibility Mask PhotoZ (VMPZ-ID) \ac{PF} \citep{DR1-TP028}. Finally, the NIR PSF is modelled by~\citet{DR1-TP005}.

\subsection{\label{ssc:imaging}EXT processing}

\Euclid is said to have a third instrument besides VIS and NISP: this is because of the EXT data,  which provide the homogeneous ground-based imaging needed to derive the photometric redshifts and \acp{SED} of the objects detected in MER. We use as input a series of large surveys. For the \ac{EWS}, the northern data are provided by \ac{UNIONS} and the southern data by \ac{DES}, as described in  \citet{UNIONS2025} and \citet{DES2021}, respectively.  \citet{DR1-TP006} details how these data are ingested in the \ac{SGS}, how they are recalibrated, and how their quality is assessed. 

While the EXT processing is improved with respect to Q1, a few issues are still present in the DR1 EXT products, as discussed in section~6 of \citet{DR1-TP006}. This is especially the case with the \ac{Pan-STARRS} imaging \citep{PanSTARRS2016}, where an incorrect handling of the flags by the SGS has led to areas with degraded photometry. In the same spirit as for the astrometric NIR issue outlined in \cref{ssc:NIR}, a dedicated {\tt HEALPix}\footnote{\url{https://healpix.sourceforge.io}} \citep{Gorski2005} information map is provided in the DR1-Foundation VMPZ-ID products \citep{DR1-TP028}. We stress here that this issue arises from our processing, not from the \ac{UNIONS} data, which are of excellent quality. The EXT pipelines have been updated so that this issue will not be present any more in the DR2 release.

\subsection{\label{ssc:mer}MER data processing}

MER is the \ac{PF} in charge of the creation of \Euclid data stacks from the individually calibrated exposures delivered by VIS and NIR on a common astrometric grid. It also carried out the detection of objects and measurement of their properties. The DR1 version of the MER PF builds on the version that was used for Q1 \citep{Q1-TP004}, with various improvements such as the computation of a local noise level, and the addition of new quantities in the catalogue. These, together with the extensive validation of the images and catalogues, are presented in \citet{DR1-TP009}.

\citet{DR1-TP010} makes use of comparison with the COSMOS2025 catalogue \citep{COSMOS2025} independently measured on one of the \acp{EAF}. This provides an extensive check of the MER photometry, both at the \ac{EWS} depth and at the greater depth that can be reached on this field.  

Note that the MER processing is normally triggered if all the VIS, NIR, and EXT data exist. In some cases, MER has been run when only the VIS and NIR data were present, in order to allow for the spectroscopic processing to run, as it needs the MER catalogue as input. For this reason, DR1-Foundation does not contain MER data for some of the data sets. The type of MER processing per data set is provided in \Cref{tab:le2mer}. `MER Wide' is the standard processing for each \ac{ROS}, while `MER Deep' combines all \acp{ROS} available for depth. `MER VIS + NIR' means that MER was triggered even if no EXT data are available, in order to be able to run the SIR pipeline. Note that in this table we break the \ac{EAF} data set into its three separate fields, since we did not have EXT processing of the ground-based data for AEGIS.

\begin{table}
\caption{\label{tab:le2mer}Types of MER processing available per data set.}
\begin{center}
    \begin{tabular}{ll}
\hline\hline
\noalign{\vskip 2pt}
    Data set &  MER Processing\\
\hline
\noalign{\vskip 2pt}
EWS & MER Wide \\
\hline
\noalign{\vskip 2pt}
EDFs & MER Deep + Wide\\
\hline
\noalign{\vskip 2pt}
EAF--COSMOS & MER Deep + Wide \\
EAF--CDFS   & MER Deep + Wide \\
EAF--AEGIS  & MER VIS + NIR \\
\hline
\noalign{\vskip 2pt}
SelfCal  & MER Wide \\
PSF &  \dots \\
F010 & MER VIS + NIR \\
F014 & MER Wide \\
ILS &  MER VIS + NIR \\
WAVES & \dots \\
SSO &  MER VIS + NIR \\
EMDS-1 & \dots  \\
EMDS-2 & \dots \\
EMDS-3 & MER VIS + NIR \\
EMDS-4 & MER VIS + NIR \\
\hline
    \end{tabular}
    \end{center}

\end{table}

\subsection{\label{ssc:phz}PHZ data processing}

The PHZ \ac{PF} provides the photometric redshifts and physical properties of the sources in the MER catalogue. The Q1 version of the PF is described in \citet{Q1-TP005}. For DR1, the main addition is the availability of \acp{EAF} to calibrate the {\tt NNPZ} \citep[see][]{EP-Enia} reference sample so that the object redshift probability distributions can be derived using this package \citep{DR1-TP011}. Improvements have also been made for the computation of the physical properties of sources, as described in \citet{DR1-TP012}. Overall, DR1-Foundation provides photometric redshifts and classification for 880 million sources over the 1900\,deg$^2$ of the \ac{EWS}. 

\subsection{\label{ssc:vmpz}VMPZ-ID data processing}

The VMPZ-ID is formally a \ac{LE3} processing function, whose purpose is to determine the selection function and create random catalogues of the imaging part of the \ac{EWS}. However, in order  to compute these final products, it also assembles information about the data drawn directly from \ac{LE2} products. Some of these products are very useful for the scientific exploitation of the DR1-Foundation data set, and are close enough to their LE2 origin that they have been included in this release. All these products are either masks or maps of useful quantities distributed as {\tt HEALPix} maps. The construction of these numerous products is described in~\citet{DR1-TP028}. The VMPZ-ID PF was only run on the \ac{EWS} and \ac{EDF} data sets.

\subsection{\label{ssc:spectro}SIR and SPE data processing}

The processing of the spectroscopic exposures is carried out in two \acp{PF}. First, the spectra are extracted by the \ac{SIR} PF. The processing is described by \citet{Q1-TP006}, with numerous improvements such as better persistence flagging and a more efficient extraction algorithm~\citep{DR1-TP023}. This was made possible by the accurate calibration of the instrument, with the spectral curvature, dispersion solution, PSF as a function of wavelength, and other parameters being accounted for~\citep{DR1-TP024}. As a result, a spectrum has been extracted in DR1 for all objects in the MER catalogue with \HE$<23.5$; this is a full magnitude deeper than what was done for Q1. 

The spectra are then passed to the SPE PF for redshift estimation, spectral classification, and spectral feature measurements. The DR1 version of the PF builds on the one used for Q1 \citep{Q1-TP007}, with improved templates and better measurement methods~\citep{DR1-TP025}. The SPE PF also benefits from the higher quality of the spectra delivered by SIR. As a result, a comparison with \ac{DESI} redshifts \citep{DESI-Collaboration:2024} finds good agreement, with a scatter half that of the Q1 sample. Finally, the DR1-Foundation data set contains observations of the EDF-N with both the blue and red grisms, allowing us to obtain a high purity sample that will be used to derive the spectroscopic selection function for the full DR1 data release.

As for Q1, because of the nature of the measurement where we extract a spectrum for all objects with \HE$<23.5$, we caution the users of the DR1-Foundation release against blindly using the redshifts: many spectra do not show any measurable features, yet have an entry in the SPE catalogue. \citet{DR1-TP025} discusses redshift reliability and the flags that can be used to assess this.

\subsection{Data volume}
The volume of data being made available in DR1-Foundation corresponds to approximately 1.7\,PB \citep{DR1-TP033}. Included are more than 120\,000 calibrated frames from \ac{VIS} and \ac{NISP}.  There are also uniformly reduced and calibrated EXT data, primarily from \ac{UNIONS} in EWS-North and \ac{DES} in the EWS-South, and other imaging data in the \acp{EDF} and \acp{EAF}.

Dr1-Foundation also contains over 170 million validated spectra in both red- and blue-grism modes extracted from the NISP dispersed images, most having spectroscopic redshift estimates. Over 60 million sources are identified as quasar candidates from photometric data.

There are more than a billion catalogued sources in DR1.
Focussing on those sources categorised as galaxies (specifically with \texttt{PHZ\_CLASSIFICATION=2}), the total number of over 900 million breaks down as follows:
\begin{itemize}
\item \ac{EWS}, 817\,846\,065;
\item \ac{EDS}, 49\,393\,051;
\item \acp{EAF}, 34\,427\,023.
\end{itemize}

\section{Technical papers associated with DR1}\label{sc:datapapers}
DR1-Foundation is documented by a comprehensive list of papers (\cref{tab:papers}) that provide the necessary context for its scientific exploitation. These papers are published along with this overview paper, discussing the updates made with respect to Q1, the currently known limitations of DR1, and upcoming improvements for DR2. 

\paragraph{\citet{DR1-TP027}} details the multiple \Euclid surveys and fields, including the \ac{EDS}, \acp{EAF}, calibration fields, \ac{EMDS}, the Euclid Ecliptic Survey, and the \ac{ILS} field, in addition to the \ac{EWS}. 
It also describes the Science Operations Centre (SOC) activities of the operational sky survey (OSS) generation and the production of operational products.

\paragraph{\citet{DR1-TP033}} describes the operation and architecture of the \Euclid \ac{SGS}, covering the role of the SOC, the Science Data Centres (SDCs), and the SGS common infrastructure, including the Euclid Archive System (EAS). 
It also discusses the processing methodology and daily processing operations. The production of Euclid \ac{DR1} involved the execution of 35 million processing jobs and 6.7 million core hours of computing.

\paragraph{\citet{DR1-TP002}} 
describes the \ac{VIS} \ac{DR1} data set totalling 42\,092 images, evaluates the performance of the \ac{VIS} processing function pipeline and discusses the global properties of the processed \ac{VIS} images.
In particular the levels of astrometric and photometric precision are quantified and shown to be within mission requirements.

\paragraph{\citet{DR1-TP003}} describes the near-infrared processing function (NIR), covering its processing algorithms, calibration, data products, and validation. NISP astrometric and photometric precision are also shown to be within requirements. In total, the \ac{DR1} processing produced 80\,695 calibrated frames of \ac{NISP} data.  

\paragraph{\citet{DR1-TP004}} provides further details of the \ac{NISP} astrometry, describing how instrumental distortions are modelled and corrected, with comparison to stellar positions from \textit{Gaia} \citep{GaiaDR3}.
 
\paragraph{\citet{DR1-TP005}} describes the modelling of the \ac{NISP} \ac{PSF}, part of the NIR processing function, investigating how this varies with time, focal-plane position, and wavelength.

\paragraph{\citet{DR1-TP024}} describes the spectroscopic calibration carried out by the spectroscopic SIR processing function. This provides validated spectra for more than 170 million sources.

\paragraph{\citet{DR1-TP038}} describes the effects of detector persistence in the \ac{NISP} instrument, including the masking strategy used in \ac{DR1}.
 
\paragraph{\citet{DR1-TP006}} gives a detailed description of the processing
and calibration steps applied to the ground-based optical imaging data sets (EXT) released in DR1. In particular, there is a description of how stacked data products are generated and an assessment of the EXT data quality.

\paragraph{\citet{DR1-TP009}} describes the MERge processing function, which detects objects in the \ac{VIS}, \ac{NISP}, and EXT imaging data,
measures their properties, and combines them into a single multi-wavelength catalogue.

\paragraph{\citet{DR1-TP010}} describes the use of existing data from the \ac{COSMOS} field \citep{COSMOS2025} to validate the \ac{DR1} photometry.

\paragraph{\citet{DR1-TP011}} describes the photometric redshift processing function (PHZ), which also provides spectral energy distribution fits for galaxies and stars.

\paragraph{\citet{DR1-TP012}} supplements the other PHZ paper by presenting the estimation of galaxy physical parameters for non-cosmological science, in particular studying the relationship between stellar mass and star-formation rate.

\paragraph{\citet{DR1-TP025}} describes the spectroscopic SPE processing function, which analyses the 1-dimensional spectra from \ac{NISP} It also carries out spectroscopic
classification, redshift determination, and the measurement of spectral features.

\paragraph{\citet{DR1-TP028}} describes the series of photometric masks developed by the VMPZ-ID processing function for use with the \ac{DR1} data.  This includes the effective coverage mask, as well as specific masks representing spatial variations of various properties of the observations.

\paragraph{\citet{DR1-TP026}} describes the overall calibration strategy, the performance of the calibration plan with respect to top-level mission calibration requirements, and current short-comings, as well as planned improvements. 

\paragraph{\citet{DR1-TP026b}} assesses the spacecraft performance in the L2 environment up until October 2026, covering -- among other issues -- stray-light sensitivity, thermo-optical properties, molecular contamination, tracking stability, and meteoroid impacts.
\\

Additional papers are still in preparation (\cref{tab:papers}). A complete set will appear in support of the full DR1 in June 2027.

\begin{table*}
\caption{\label{tab:papers}List of papers supporting DR1.}
\begin{center}
    \begin{tabular}{lll}
\hline\hline
\noalign{\vskip 2pt}
    No. & Title & Authors \\
\hline
\noalign{\vskip 2pt}
\po1 & Data release overview & Aussel et al. \\
\po2 & The \Euclid surveys at DR1 & Tereno et al. \\
\po3 & An overview of the Science Ground Segment architecture and operations & Ruffenach et al. \\
\po4& VIS processing and data products & McCracken et al.\\
\po5 & Near-infrared processing and data products & Polenta et al.\\
\po6 & NIR astrometric calibration & Radovich et al. \\
\po7 & NISP-P point-spread function modelling pipelines & da Silva et al. \\
\po8 & SIR spectroscopic calibration & Fumana et al. \\
\po9 & Persistence in NISP detectors & Kubik et al. \\
10 & Colouring \Euclid with external ground-based imaging data & Corcho-Caballero et al. \\
11 & The MERge processing & Kümmel et al. \\
12 & Multi-band photometry validation on COSMOS extragalactic sources & Gasparetto et al. \\
13 & Photometric redshift pipeline overview and validation & Hartley et al. \\
14 & The PHZ processing function for the physical properties of galaxies & Enia et al. \\
15 & Performance of the redshift measurement and spectroscopic analysis & Le Brun et al. \\
16 & \Euclid photometric masks & Odier et al. \\
17 & Calibration strategy and observations & Schirmer et al. \\
18 & Spacecraft performance in the L2 environment & Schirmer et al. \\
\hline
\noalign{\vskip 2pt}
  & The SIR spectroscopic processing function& in prep.\\
  & Euclid Archive System: data processing system and data storage system & in prep.\\ 
  & The Euclid Data Space & in prep.\\
  & \textit{Gaia}-based calibration & in prep.\\
  & EXT Dark Energy Survey data reduction & in prep.\\
  & EXT PSF modelling pipeline & in prep.\\
  & $N(z)$ estimation and calibration performance & in prep.\\
  & $N(z)$ estimates validation& in prep.\\
  & Shear overview, requirements, methods, and validation& in prep.\\
  & Shear Data Catalogue & in prep.\\
  & Shear calibration& in prep.\\
  & MetaCal & in prep.\\
  & \Euclid VIS PSF model& in prep.\\
  & \Euclid VIS PSF calibration & in prep.\\
  & \Euclid VIS PSF field calibration & in prep.\\
  & Thermal property tracking in the image domain & in prep.\\
  & Mitigation strategies for the photometric 2-pt correlation function& in prep.\\
  & Photometric visibility mask and randoms for the 2-pt correlation function& in prep.\\
  & Photometric masks and effective coverage for cluster detection& in prep.\\
  & Photometric masks and combined catalogues& in prep.\\
  & In-orbit performance of the NISP Instrument& in prep.\\
\noalign{\vskip 2pt}
\hline
     \end{tabular}
\end{center}
\end{table*}

\section{Conclusions}
Covering around 2000\,deg$^2$, the publicly available Euclid DR1-Foundation is, to date, the largest sky area observed from space at optical and near-infrared wavelengths. The sub-arcsecond resolution imaging is complemented by slitless spectroscopy. These high-quality data have begun to help answer a vast number of scientific questions -- some of which had never been conceived of when \Euclid was designed -- and enable numerous new discoveries. 

Compared with earlier modest releases, science exploitation for DR1-Foundation is proceeding differently, now that the data sets are so extensive. The Q1 data release was accompanied by a large number of papers presenting discoveries based on those data. For DR1-Foundation, such papers will appear when they are ready, some along with DR1-Foundation, others in the very near future, but many more in the coming months. These tackle a wide range of different science cases that can be addressed with the DR1-Foundation data, from the search for solar-system objects to the discovery of distant quasars and spectacular strong lenses.

In June 2027, DR1 will be completed with shear measurements, along with weak-lensing and galaxy-clustering correlation functions, thereby beginning to deliver the cosmological science for which \Euclid was built. We stress that the upcoming full DR1 release will use the DR1-Foundation products. No product released as part of DR1-Foundation will be reprocessed for DR1. DR1 will only add the SHE and numerous LE3 products. 

\Euclid continues to operate smoothly and with extraordinarily high reliability. At the time of writing, \Euclid has observed approximately 8700\,deg$^2$ of the sky. Together with DR1, these data are being used to continuously improve the pipelines for DR2, scheduled for 2029.

%
%

\begin{acknowledgements}
\AckEC
\AckDRone
\end{acknowledgements}

%
%

\bibliography{Euclid,dr1overview} 
%
%

%

  
%

\end{document}